\documentclass[10pt,conference,letterpaper]{IEEEtran}
\IEEEoverridecommandlockouts
\usepackage{cite}
\usepackage{amsmath,amssymb,amsfonts}
\usepackage{graphicx}
\usepackage{textcomp}
\usepackage{xcolor}
\usepackage{booktabs}
\usepackage{array}
\usepackage{multirow}
\usepackage{ragged2e}
\usepackage{microtype}
\usepackage{comment}
\usepackage{needspace}
\usepackage{orcidlink}
\usepackage{tikz}
\usetikzlibrary{positioning,fit,calc,arrows.meta,decorations.pathreplacing}
\def\BibTeX{{\rm B\kern-.05em{\sc i\kern-.025em b}\kern-.08em
T\kern-.1667em\lower.7ex\hbox{E}\kern-.125emX}}
\begin{document}

\title{When Discovery Becomes a Storm:  A ROS~2 Discovery Model for Wireless Robotic Networks
}

\author{
  \IEEEauthorblockN{%
    Yeonwoo~Choi\,\textsuperscript{\textdagger}\, \orcidlink{0009-0007-1437-0896},
    Sanghoon~Lee\,\textsuperscript{\textdagger}\, \orcidlink{0000-0002-8160-8952},
    and Kyung-Joon~Park\,\textsuperscript{*}\ \orcidlink{0000-0003-4807-6461}}
  \IEEEauthorblockA{Department of Electrical Engineering and Computer Science, DGIST, Daegu, Republic of Korea\\
  Email: \{cyw040306, leesh2913, kjp\}@dgist.ac.kr}}

\maketitle

\begingroup

\renewcommand{\thefootnote}{\textdagger}
\footnotetext{Both authors contributed equally to this research.}

\renewcommand{\thefootnote}{*}
\footnotetext{Corresponding author}

\endgroup

\begin{abstract}
In Robot Operating System~2 (ROS~2), Data Distribution Service (DDS) participants must discover one another before exchanging data.
In wireless environments, delayed or lost discovery messages cause reliability timers to expire, triggering retransmissions that intensify channel contention and further delay the delivery of discovery messages.
This self-reinforcing feedback can escalate into a \emph{discovery storm}.
Existing models characterize discovery demand under fixed delivery conditions, but do not capture how shared-channel delay changes protocol state and generates further traffic.
To address this issue, we present the first closed-loop analytical model of ROS~2 discovery that characterizes how delay-induced feedback amplifies retransmission overhead and leads to severe discovery storms.
Our model represents channel contention as a shared service process, coupling message-delivery latency with receiver states and reliability timers.
The model predicts both discovery completion time and per-class message counts.
We validate the model through 1,350 experimental runs across 90 topology configurations.
An open-loop airtime baseline captures only a fraction of the high-load completion time.
The closed-loop model reproduces this rise and conservatively upper-bounds the observed high-load range.
Guided by insights from the model, we further design a response-aware discovery policy that reduces mean discovery completion time by 25.3\% to 39.7\%.
\end{abstract}


\section{Introduction}

Wireless robotic networks coordinate robot functions across distributed processes.
These systems commonly rely on the Robot Operating System~2~(ROS~2), a de facto standard software framework that uses the Data Distribution Service~(DDS) for communication~\cite{macenski2022robot}.
Before application data can flow, these processes must discover one another and exchange the communication metadata needed to establish data paths~\cite{OMG-RTPS25}.
This decentralized startup supports modular deployment and dynamic joining, but the required metadata must be delivered and recognized before the expected communication paths become ready.
In a shared wireless environment, discovery messages contend for the channel and can therefore delay system startup or prevent it from completing.

A \emph{discovery storm} is a startup regime in which discovery demand exhausts the available wireless service, causing discovery to complete late or not at all~\cite{lee2026discovery}. 
Even two networked devices, such as a robot computer and a server, can reach this regime when many data publishers and subscribers start discovery together.
By consuming the available wireless bandwidth with redundant retransmission traffic, a discovery storm paralyzes network startup and prevents distributed processes from exchanging time-critical application data.
Therefore, ensuring reliable ROS~2 communications requires a comprehensive closed-loop model to capture the self-reinforcing dynamics of ROS~2 DDS discovery.

Prior work examines discovery overhead from three complementary perspectives.
A deterministic airtime analysis quantifies the deployment-driven initial discovery burst and identifies when it approaches channel capacity~\cite{lee2026discovery}.
Analytical DDS latency models estimate reliability recovery, but rely on an externally fixed packet-delivery ratio~\cite{park2025analytical}.
Other studies focus on reducing initial discovery demand via Bloom filters~\cite{sanchez2011bloom}, content-based filtering~\cite{an2014content}, centralized discovery~\cite{luo2025centralized}, or node composition~\cite{macenski2023impact}.
The remaining gap is the causal link through which shared-channel delay postpones receiver recognition, prompting further delivery-check and repair traffic.
Modeling this link explains why identical first-transmission workloads can yield different repair counts and completion times.

In this paper, our contributions are as follows:
\begin{itemize}
\item \emph{Modeling.} We formalize the closed-loop dynamics of the discovery storm, in which shared-channel delay triggers periodic timer-driven repair that reenters the shared channel and deepens the delay, separating this repair from the topology-dependent first-transmission burst and quantifying its amplification.

\item \emph{Validation.} In the 18 high-load configurations, an open-loop airtime baseline without the feedback predicts only about 17\% of the observed high-load completion time.
The closed-loop model captures this rise, upper-bounding roughly 90\% of high-load completion-time and message observations as a conservative upper-range estimate.

\item \emph{Mitigation.} The model isolates the periodic reliability messages as a tunable source of contention, and we pace them to receiver responses so that they do not contend with pending discovery traffic, reducing participant-wise mean discovery completion time by \(25.3\%\) to \(39.7\%\).
\end{itemize}

The remainder of this paper is organized as follows.
Section~II reviews DDS discovery and the delay-induced repair amplification that motivates the model.
Section~III develops the closed-loop discovery model.
Section~IV evaluates the model and the response-aware policy on an 802.11n testbed.
Section~V discusses related work, and Section~VI concludes.

\section{Background and Motivation}
\label{sec:background-motivation}

\subsection{DDS Discovery Protocol}
\label{subsec:dds-discovery-protocol}
\label{subsec:discovery-completion}

\begin{figure}[!t]
  \centering
  \resizebox{0.88\columnwidth}{!}{%
    \begin{tikzpicture}[
  font=\normalsize,
  host/.style={draw=black!70,line width=0.75pt,rounded corners=1.2pt,fill=black!2,minimum width=52mm,minimum height=52mm,inner sep=0pt},
  participant/.style={draw=black!55,line width=0.55pt,rounded corners=0.8pt,fill=white,minimum width=22mm,minimum height=34mm,inner sep=0pt},
  endpoint/.style={draw=black!45,line width=0.4pt,rounded corners=0.5pt,fill=black!5,minimum width=18.5mm,minimum height=4.6mm,inner xsep=1.5pt,inner ysep=0.8pt,font=\footnotesize},
  hosttitle/.style={font=\normalsize,anchor=north west},
  plabel/.style={font=\footnotesize,anchor=north west,align=left},
  annot/.style={font=\small,inner sep=1pt},
  dots/.style={font=\normalsize,inner sep=0pt}
]
  \node[host] (h1) at (0,-3mm) {};
  \node[host] (hN) at (59mm,-3mm) {};
  \node[dots] at ($(h1.east)!0.5!(hN.west)$) {$\cdots$};
  \node[hosttitle] at ($(h1.north west)+(2mm,-0.05mm)$) {Host $1$};
  \node[hosttitle] at ($(hN.north west)+(2mm,-0.05mm)$) {Host $N_H$};

  \begin{scope}[shift={($(h1.center)+(-13.5mm,2mm)$)}]
    \node[participant] (p11) at (0,0) {};
    \node[plabel] at ($(p11.north west)+(1mm,-0.6mm)$) {participant\\process $1$};
    \node[endpoint] (p11e1) at (0,4.65mm) {endpoint 1};
    \node[endpoint] (p11e2) at (0,-0.75mm) {endpoint 2};
    \node[font=\footnotesize] at (0,-5.45mm) {$\vdots$};
    \node[endpoint] (p11eN) at (0,-12.35mm) {endpoint $N_E$};
  \end{scope}

  \begin{scope}[shift={($(h1.center)+(13.5mm,2mm)$)}]
    \node[participant] (p1P) at (0,0) {};
    \node[plabel] at ($(p1P.north west)+(1mm,-0.6mm)$) {participant\\process $N_P$};
    \node[endpoint] (p1Pe1) at (0,4.65mm) {endpoint 1};
    \node[endpoint] (p1Pe2) at (0,-0.75mm) {endpoint 2};
    \node[font=\footnotesize] at (0,-5.45mm) {$\vdots$};
    \node[endpoint] (p1PeN) at (0,-12.35mm) {endpoint $N_E$};
  \end{scope}
  \node[font=\small] at ($(p11.east)!0.5!(p1P.west)$) {$\cdots$};

  \begin{scope}[shift={($(hN.center)+(-13.5mm,2mm)$)}]
    \node[participant] (pN1) at (0,0) {};
    \node[plabel] at ($(pN1.north west)+(1mm,-0.6mm)$) {participant\\process $1$};
    \node[endpoint] at (0,4.65mm) {endpoint 1};
    \node[endpoint] at (0,-0.75mm) {endpoint 2};
    \node[font=\footnotesize] at (0,-5.45mm) {$\vdots$};
    \node[endpoint] at (0,-12.35mm) {endpoint $N_E$};
  \end{scope}

  \begin{scope}[shift={($(hN.center)+(13.5mm,2mm)$)}]
    \node[participant] (pNP) at (0,0) {};
    \node[plabel] at ($(pNP.north west)+(1mm,-0.6mm)$) {participant\\process $N_P$};
    \node[endpoint] at (0,4.65mm) {endpoint 1};
    \node[endpoint] at (0,-0.75mm) {endpoint 2};
    \node[font=\footnotesize] at (0,-5.45mm) {$\vdots$};
    \node[endpoint] at (0,-12.35mm) {endpoint $N_E$};
  \end{scope}
  \node[font=\small] at ($(pN1.east)!0.5!(pNP.west)$) {$\cdots$};

  \draw[decorate,decoration={brace,mirror,amplitude=2.5pt}]
    ($(p11.south west)+(0,-1.7mm)$) -- ($(p1P.south east)+(0,-1.7mm)$)
    node[midway,below=1.5mm,annot] {$N_P$ participant processes};
  \draw[decorate,decoration={brace,mirror,amplitude=2.5pt}]
    ($(pN1.south west)+(0,-1.7mm)$) -- ($(pNP.south east)+(0,-1.7mm)$)
    node[midway,below=1.5mm,annot] {$N_P$ participant processes};

  \draw[decorate,decoration={brace,mirror,amplitude=3pt}]
    ($(h1.south west)+(0,-3.4mm)$) -- ($(hN.south east)+(0,-3.4mm)$)
    node[midway,below=1.8mm,annot] {$N_H$ wireless hosts};
\end{tikzpicture}%
  }
  \caption{Homogeneous DDS discovery topology used in the evaluation,
  with \(N_H\) wireless hosts, \(N_P\) participants per host, and \(N_E\)
  user endpoints per participant.}
  \label{fig:discovery-topology}
  \vspace{-1em}
\end{figure}
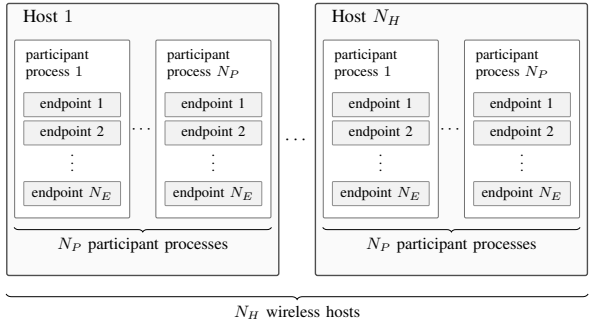

A DDS deployment consists of hosts, participants, and endpoints.
A host runs software processes that create DDS participants, and each participant contains endpoints that publish or subscribe to application data, such as sensor readings or motion commands.
In general, hosts may contain different numbers of participants, and participants may advertise different numbers of endpoints.
Fig.~\ref{fig:discovery-topology} illustrates the homogeneous special case used for controlled evaluation, with \(N_H\) hosts, \(N_P\) participants per host, and \(N_E\) user endpoints per participant.
Communication within one host does not traverse the shared wireless link.
Two hosts are therefore the smallest setting that exposes cross-host discovery over the shared channel.

Discovery proceeds in two stages.
The Simple Participant Discovery Protocol~(SPDP) first announces each participant's presence and network addresses.
After remote participants recognize one another, the Simple Endpoint Discovery Protocol~(SEDP) exchanges metadata describing their publishing and subscribing endpoints.
SEDP carries this metadata in Real-Time Publish-Subscribe~(RTPS) DATA submessages.
One RTPS packet carried over the User Datagram Protocol~(UDP) may contain several submessages~\cite{OMG-RTPS25}.

SEDP provides reliable delivery through a check-and-repair cycle.
A sender transmits SEDP DATA and uses a HEARTBEAT to announce the available DATA sequence numbers. {Sequence numbers label SEDP DATA items so that a receiver can identify which items are missing.} The receiver returns an ACKNACK identifying which DATA has arrived and which remains missing, and the sender retransmits the requested metadata.
The first transmission of an endpoint's metadata to a receiver is Initial SEDP DATA.
Each later transmission of the same metadata to that receiver is Repair SEDP DATA~\cite{OMG-RTPS25}.

Discovery completion depends on the receiver state rather than sender transmission.
Discovery is completed only after every receiver has processed the metadata for all expected remote endpoints.
DATA that has been generated but remains queued or in service is still absent from the receiver state. {Here, a reliability timer schedules a protocol decision that checks or responds to the receiver's delivery state.} If a reliability timer expires before that DATA is processed, the protocol treats the delayed metadata as missing and may initiate the same repair path as an actual loss.

\subsection{Delay and Repair Amplification}
\label{subsec:delay-amplification}

Prior studies show that DDS performance depends on multicast load and wireless-channel conditions~\cite{peeroo2022exploring,almadani2015performance}.
Wireless loss can trigger repair, yet loopback runs produce Repair SEDP DATA despite negligible network loss (Subsection~\ref{subsec:completion-scaling}), showing that loss is not required.
We therefore focus on the causal path in which delayed delivery postpones receiver recognition and activates the reliability procedure.
The resulting checks, replies, and retransmissions increase contention and further delay the remaining metadata.

Comparing wired and wireless discovery illustrates this amplification.
The wired environment provides a low-delay reference, while the wireless environment allows discovery messages to be delayed or remain unreceived when a receiver checks what has arrived.
We use a representative high-load deployment with two hosts, five participants per host, and twenty user endpoints per participant, namely \(N_H=2\), \(N_P=5\), and \(N_E=20\).
All processes start simultaneously and advertise the same participants and endpoints under both conditions.
Table~\ref{tab:repair-amplification-example} separates SPDP announcements, initial and repair SEDP DATA, and the associated reliability-control messages.

\begin{table}[!t]
\caption{Discovery message counts averaged over 15 runs for
\(N_H=2\), \(N_P=5\), and \(N_E=20\)}
\label{tab:repair-amplification-example}
\vspace{-1em}
\centering
\scriptsize
\setlength{\tabcolsep}{2.5pt}
\begin{tabular}{@{}>{\raggedright\arraybackslash}m{0.31\columnwidth}
                >{\raggedleft\arraybackslash}m{0.19\columnwidth}
                >{\raggedleft\arraybackslash}m{0.22\columnwidth}
                >{\raggedleft\arraybackslash}m{0.16\columnwidth}@{}}
\toprule
{\scriptsize\bfseries Message category} &
{\scriptsize\bfseries Wired} &
{\scriptsize\bfseries Wireless} &
{\scriptsize\bfseries \shortstack{Wireless/\\Wired ratio}} \\
\midrule
SPDP & 327.3 & 529.8 & 1.6 \\
Initial SEDP DATA & 1,850.0 & 1,850.0 & 1.0 \\
Repair SEDP DATA & 0.07 & 3,644.0 & 54,660 \\
HEARTBEAT & 102.5 & 14,036.1 & 137.0 \\
ACKNACK & 100.4 & 3,798.1 & 37.8 \\
\midrule
\textbf{Total} & \textbf{2,380.2} & \textbf{23,858.1} & \textbf{10.0} \\
\bottomrule
\end{tabular}
\vspace{-0.5em}
\end{table}

\begin{figure}[!t]
  \centering
  \includegraphics[width=\columnwidth,height=0.72\columnwidth,keepaspectratio,trim=0 0 0 12pt,clip]{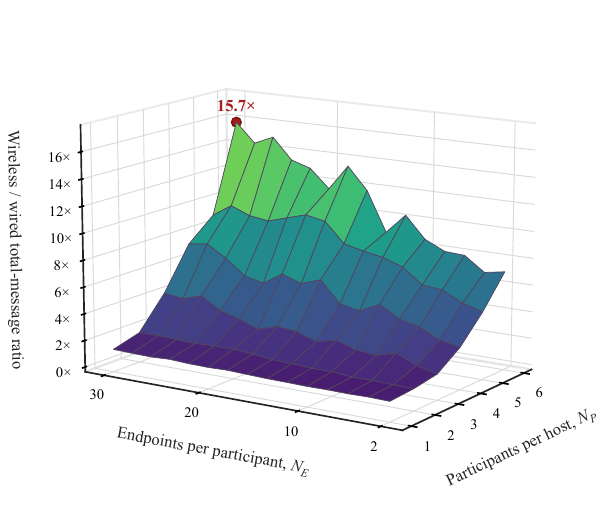}
  \vspace{-0.5em}
  \caption{Ratio of the wireless and wired run-mean total
  discovery-message counts at each evaluated system size. Color follows the
  surface height.}
  \label{fig:traffic-ratio-surface}
  \vspace{-1em}
\end{figure}

Across the measured runs, both environments produce an average of 1,850 initial SEDP DATA submessages.
This confirms that the large difference in total traffic does not come from advertising more endpoints.
Participant announcements, delivery checks, receiver replies, and retransmissions produce only 2,380.2 discovery submessages in the wired environment.
In the wireless environment, the same categories raise the total to 23,858.1~submessages, which is 10~times more than that of the wired traffic.
Fig.~\ref{fig:traffic-ratio-surface} extends this representative comparison across all evaluated configurations.
The wireless-to-wired ratio of run-mean total discovery-message counts rises jointly with \(N_P\) and \(N_E\) and reaches 15.7 in the most~amplified~\mbox{configuration.}

\section{Closed-loop discovery model}
\label{sec:model}

\looseness=-1 {This section turns the feedback mechanism described in Subsection~\ref{subsec:delay-amplification} into a closed-loop model.
The deployment input describes the system structure, a DDS profile supplies protocol behavior, and a wireless-service profile describes how transmissions are served.
The protocol component generates messages from receiver state and timers.
The wireless-service component determines when each message leaves the sender and reaches the receiver.
Each completed delivery updates the state used by subsequent protocol decisions.
A different wireless network is represented by its own wireless-service profile.}

\subsection{Structural Discovery Workload}
\label{subsec:model-protocol-structure}
\label{subsec:structural-workload}
\label{subsec:first-transmission-reference}

Let \(\mathcal H=\{0,\ldots,N_H-1\}\) be the wireless hosts, \(\mathcal P=\{0,\ldots,n-1\}\) the DDS participants, and {\(h:\mathcal P\rightarrow\mathcal H\) maps each participant to the host on which it runs.} {One descriptor is the SEDP metadata that advertises one publishing or subscribing endpoint.} Participant \(i\) advertises \(D_{i,s}\) descriptors of SEDP metadata type \(s\in\{\mathrm{pub},\mathrm{sub}\}\).
Each participant owns separate built-in discovery streams for publication and subscription metadata.
For a directed cross-host pair \((i,j)\), \(i\)'s publication writer sends {descriptions of its data-sending endpoints} to \(j\)'s publication reader, while its subscription writer sends {descriptions of its data-receiving endpoints} to \(j\)'s subscription reader.
Each stream maintains independent reliability state and timers.
Consequently, every directed pair contributes {one reliable publication relation and one reliable subscription relation.}

\begin{table}[!t]
\caption{Notation for the Closed-Loop Discovery Model}
\label{tab:model-notation}
\centering
\scriptsize
\setlength{\tabcolsep}{3pt}
\renewcommand{\arraystretch}{1.04}
\begin{tabular}{@{}>{\centering\arraybackslash}m{0.27\columnwidth}
                >{\RaggedRight\arraybackslash}m{0.65\columnwidth}@{}}
\toprule
\textbf{Symbol} & \textbf{Meaning} \\
\midrule
\(N_H,n,N_P,N_E\) &
Hosts, total participants, participants per host, and user endpoints per
participant \\
\midrule
\(\mathcal H,\mathcal P,h\) &
Host set, participant set, and the map assigning each participant to a host \\
\midrule
\(\mathcal R,\ i,j,s,d\) &
Directed cross-host participant pairs, with sender, receiver, metadata-type,
and descriptor indices \\
\midrule
\(D_{i,s},F,N_{\mathrm{rel}}\) &
Type-\(s\) descriptors advertised by \(i\), required descriptor deliveries,
and reliable SEDP relations \\
\midrule
\(\begin{array}{c}
\alpha_{\mathrm{HB}},\alpha_{\mathrm{ACK}}\\
\alpha_e,\alpha_m,\alpha_u
\end{array}\) &
Profile counts used for reference HEARTBEAT, ACKNACK, and SPDP traffic \\
\midrule
\(v,c\) &
Generated RTPS/UDP packet and one uplink or downlink fragment transmission \\
\midrule
\(B(c),\kappa(c),a(c)\) &
Fragment byte count, unicast or multicast mode, and wireless service demand \\
\midrule
\(W_i,I_i(t)\) &
Source-queue capacity and admitted bytes of participant \(i\) \\
\midrule
\(r(c),b(c),f(c)\) &
Service-request, service-start, and service-finish times of fragment \(c\) \\
\midrule
\(P_{ij},R^{(s)}_{ij},X^{(s)}_{ij,d}\) &
Participant recognition, SEDP relation, and receiver descriptor states \\
\midrule
\(C^{(s)}_{ij,d}\) &
Sender acknowledgment and repair state for one descriptor \\
\midrule
\(\mathsf E_n,\Delta\mathsf E_n,(e,t),\prec\) &
Pending events, newly scheduled events, a timed event, and event ordering \\
\midrule
\(\tau_A,\tau_N,\tau_S,\tau_H,\tau_P\) &
HEARTBEAT-response, repair-response, suppression, HEARTBEAT, and SPDP
intervals \\
\midrule
\(\begin{array}{c}
T_{\mathrm{disc}}\\
M_{\mathrm{init}},M_{\mathrm{repair}}\\
M_{\mathrm{HB}},M_{\mathrm{ACK}}\\
M_{\mathrm{SPDP,m}}\\
M_{\mathrm{SPDP,u}}
\end{array}\) &
\parbox[c][5\baselineskip][c]{\linewidth}{\RaggedRight
Discovery completion time and generated counts by message category} \\
\bottomrule
\end{tabular}
\vspace{-0.5em}
\end{table}

{Table~\ref{tab:model-notation} summarizes the notation used throughout the closed-loop discovery model.
We denote the set of directed cross-host pairs by \(\mathcal R\), the total number of required descriptor deliveries across these relations by \(F\), and the number of reliable SEDP relations by \(N_{\mathrm{rel}}\):}
\begin{equation}
\begin{aligned}
\mathcal R
  &=\{(i,j)\in\mathcal P^2:h(i)\ne h(j)\},\\
F &=\sum_{(i,j)\in\mathcal R}
     \sum_{s\in\{\mathrm{pub},\mathrm{sub}\}}D_{i,s},\\
N_{\mathrm{rel}}&=2|\mathcal R|.
\end{aligned}
\label{eq:f-and-q}
\end{equation}

The formulation permits uneven participant placement and participant-specific descriptor counts under a common wireless-service and DDS implementation profile.
In the homogeneous specialization used for evaluation, \(D_{i,s}=D_s\) and
\begin{equation}
\begin{aligned}
F&=N_H(N_H-1)N_P^2
   (D_{\mathrm{pub}}+D_{\mathrm{sub}}),\\
N_{\mathrm{rel}}&=2N_H(N_H-1)N_P^2 .
\end{aligned}
\label{eq:homogeneous-workload}
\end{equation}

Even with \(N_H=2\), \(F\) grows quadratically with \(N_P\) and in proportion to the number of advertised endpoints.

The failure-free reference delivers every required descriptor on its first transmission.
It therefore has
\begin{equation}
M^{0}_{\mathrm{init}}=F,\qquad M^{0}_{\mathrm{repair}}=0 .
\label{eq:wired-sedp-baseline}
\end{equation}
{ Let \(\alpha_{\mathrm{HB}}\) and \(\alpha_{\mathrm{ACK}}\) be the HEARTBEAT and ACKNACK counts per reliable SEDP relation.
For SPDP, let \(\alpha_e\) be the announcement events per participant, \(\alpha_m\) the multicast SPDP count per event, and \(\alpha_u\) the average unicast SPDP count per directed cross-host pair.
The remaining reference counts are
\begin{equation}
\begin{aligned}
M^{0}_{\mathrm{HB}}&=\alpha_{\mathrm{HB}}N_{\mathrm{rel}},&
M^{0}_{\mathrm{ACK}}&=\alpha_{\mathrm{ACK}}N_{\mathrm{rel}},\\
M^{0}_{\mathrm{SPDP,m}}&=\alpha_e\alpha_m|\mathcal P|,&
M^{0}_{\mathrm{SPDP,u}}&=\alpha_u|\mathcal R|.
\end{aligned}
\label{eq:wired-profile-parameters}
\end{equation}
} It serves as the counterfactual against which timer-generated delivery checks and repair traffic are measured.
\subsection{Shared Wireless Service}
\label{subsec:shared-medium}

\looseness=-1 {Let \(v\) denote one generated RTPS/UDP packet, which may contain multiple discovery submessages.
Packet \(v\) can be divided into IP fragments when required by the configured maximum transmission unit~(MTU).
Each fragment then follows the configured transmission route, the ordered path from sender to receiver.
In the evaluated profile, this path is an uplink from the sender followed by a downlink to the receiver, and the model represents the two transmissions as separate service stages.} {Let \(c\) denote one uplink or downlink fragment transmission waiting for channel service.
For such \(c\), \(B(c)\) is the transmitted byte count and \(\kappa(c)\in\{\mathrm{u},\mathrm{m}\}\) indicates whether the transmission is unicast or multicast.} Its wireless service demand is
\begin{equation}
a(c)=O_{\kappa(c)}+\frac{8B(c)}{R_{\kappa(c)}} ,
\label{eq:airtime-functions}
\end{equation}
where \(R_{\kappa(c)}\) is the configured physical-layer rate and \(O_{\kappa(c)}\) includes fixed channel-access and protocol costs.

{A generated packet first waits in the modeled participant backlog.
The model moves it to the abstract finite source queue once that queue has sufficient free capacity in bytes.} {Only admitted fragments can request channel service.
All uplink and downlink fragments then share one modeled wireless channel.
The channel carries one fragment at a time, so every other admitted fragment waits until the channel becomes available.
The model selects the fragment with the earliest service-request time.
After the access point (AP) receives one fragment of a packet, it may be ready to forward that fragment just as the sender is ready to send the next fragment.
A fixed rule chooses which transmission goes first, and Subsection~\ref{subsec:model-fit-by-class} evaluates three forms of this rule.} A selected fragment occupies the channel for \(a(c)\). {Completion of packet \(v\)'s final uplink fragment frees the byte capacity occupied by \(v\) in the modeled source queue, while complete downlink reception can update discovery state.}

{More precisely, index the fragments as \(c_1,c_2,\ldots\) in the order selected for channel service.
The symbols \(r(c_n)\), \(b(c_n)\), \(f(c_n)\) represent its service-request, service-start, and service-finish times, respectively.
For \(n\geq2\),}
\begin{equation}
\begin{aligned}
b(c_n)&=\max\{r(c_n),f(c_{n-1})\},\\
f(c_n)&=b(c_n)+a(c_n).
\end{aligned}
\label{eq:fragment-service}
\end{equation}
{The first fragment uses the channel-availability time in place of \(f(c_{n-1})\).} A downstream fragment becomes eligible only after its preceding route stage finishes.
Thus, an infrastructure-link transmission is represented by an uplink service followed by the required downlink service, while another wireless technology can supply a different ordered route.

For participant \(i\), let \(W_i\) be the finite source-queue capacity and \(I_i(t)\) its admitted bytes.
Generated {packets} wait in a first-in, first-out {participant} backlog until their byte demand fits within \(W_i-I_i(t)\).
The demand is released when the final source-side fragment completes.
This admission rule preserves traffic that temporarily exceeds the source queue as delay instead of prematurely declaring it delivered or discarded.
\looseness=-1 For {packet} \(v\), the model therefore distinguishes its generation time, source-completion time, and receiver-completion time. {Packet \(v\) reaches receiver completion only after all of its fragment-transmission stages have completed. Consequently, receiver state remains unchanged while a packet is queued or only partially transmitted.}

\begin{figure}[!t]
  \centering
  \begin{tikzpicture}[
    node distance=2.0mm,
    box/.style={
      draw=black!80,
      line width=0.5pt,
      rounded corners=1pt,
      fill=black!3,
      text=black,
      text width=0.84\columnwidth,
      minimum height=7.2mm,
      align=center,
      inner xsep=2mm,
      inner ysep=1.2mm,
      font=\scriptsize
    },
    flow/.style={
      -{Latex[length=1.5mm]},
      draw=black!80,
      line width=0.5pt
    },
    note/.style={font=\scriptsize,text=black,align=center,inner sep=0.5pt}
  ]
    \node[box] (gen) {\textbf{Discovery-message generation}\\
      Initial DATA, delivery checks, or Repair SEDP DATA};
    \node[box,below=of gen] (queue) {\textbf{Participant backlog and finite source queue}\\
      Traffic in the source queue or being transmitted remains invisible to the receiver until reception is complete};
    \node[box,below=of queue] (medium) {\textbf{Shared wireless service}\\
      Uplink and downlink fragments compete for \(a(c)\)};
    \node[box,below=of medium] (rx) {\textbf{Receiver state and reliability timers}\\
      Complete delivery updates receiver recognition. Timer expiry observes pending metadata};
    \draw[flow] (gen) -- (queue);
    \draw[flow] (queue) -- (medium);
    \draw[flow] (medium) -- (rx);
    \draw[flow,rounded corners=2pt] (rx.east) -- ++(4mm,0) |- (gen.east);
    \node[note,rotate=-90] at
      ($(rx.east)!0.5!(gen.east)+(6.5mm,-1mm)$) {checks and repairs};
  \end{tikzpicture}
  \caption{{Closed-Loop discovery service.} Channel service controls
  receiver recognition, while reliability timers return generated checks and
  repairs to the same service.}
  \label{fig:closed-loop-service}
  \label{fig:queue-abstraction}
  \vspace{-0.5em}
\end{figure}
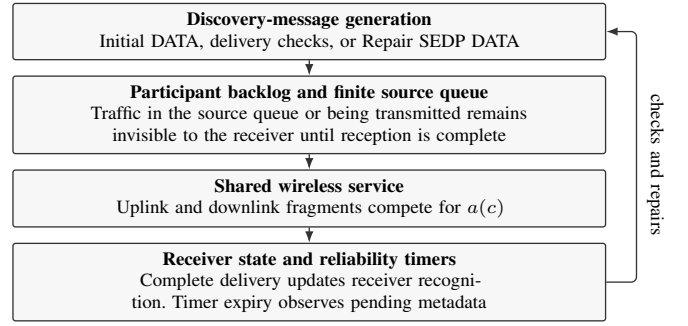

\subsection{Receiver State and Closed-Loop Evolution}

{For each directed participant pair \((i,j)\), the receiver states are} {
\begin{equation}
P_{ij}(t),\quad R^{(s)}_{ij}(t),\quad
X^{(s)}_{ij,d}(t)\in\{0,1\}.
\label{eq:receiver-state}
\end{equation}
} {The value \(P_{ij}(t)=1\) means that participant \(j\) has processed participant \(i\)'s SPDP announcement.
\(R^{(s)}_{ij}(t)=1\) means that their type-\(s\) SEDP relation is open, and \(X^{(s)}_{ij,d}(t)=1\) means that \(j\) has processed the metadata describing \(i\)'s type-\(s\) endpoint indexed by \(d\). {An open relation means that the corresponding SEDP writer and reader can begin exchanging discovery DATA and control messages.} All three states start at zero.
We omit \((t)\) when the current event time is clear.
The corresponding sender state is} {
\begin{equation}
C^{(s)}_{ij,d}(t)\in
\{\mathrm U,\mathrm R,\mathrm Q,\mathrm S,\mathrm A\}.
\label{eq:compact-state}
\end{equation}
} {Sender state \(C^{(s)}_{ij,d}\) distinguishes an unacknowledged descriptor (\(\mathrm U\)), a repair request (\(\mathrm R\)), queued repair generation (\(\mathrm Q\)), sent Repair SEDP DATA (\(\mathrm S\)), and acknowledgment (\(\mathrm A\)).} {After sending Repair SEDP DATA, the sender waits before it may generate another repair for the same descriptor.
This waiting period is called suppression.}

\begin{figure}[!t]
  \centering
  \resizebox{\columnwidth}{!}{%
  \begin{tikzpicture}[
    x=1mm,
    y=1mm,
    state/.style={
      draw=black!80,
      line width=0.5pt,
      fill=black!3,
      text=black,
      circle,
      minimum size=7.5mm,
      inner sep=0pt,
      font=\scriptsize\bfseries
    },
    finalstate/.style={state,fill=black!9},
    flow/.style={
      -{Latex[length=1.4mm]},
      draw=black!80,
      line width=0.5pt
    },
    confirm/.style={
      -{Latex[length=1.4mm]},
      draw=black!75,
      densely dashed,
      line width=0.5pt
    },
    initdot/.style={
      circle,
      fill=black,
      minimum size=1.2mm,
      inner sep=0pt
    },
    note/.style={font=\scriptsize,text=black,align=center,inner sep=0.5pt}
  ]
    \node[state] (U) at (0,0) {\(\mathrm U\)};
    \node[state] (R) at (24.5,0) {\(\mathrm R\)};
    \node[state] (Q) at (49,0) {\(\mathrm Q\)};
    \node[state] (S) at (73.5,0) {\(\mathrm S\)};
    \node[finalstate] (A) at (36.75,20) {\(\mathrm A\)};
    \node[draw=black!65,densely dashed,rounded corners=1pt,
      fit=(U)(R)(Q)(S),inner xsep=2.5mm,inner ysep=1.5mm] (nonA) {};
    \node[initdot] (init) at ($(U.west)+(-6mm,0)$) {};
    \node[above=1.0mm,note] at (init) {init};

    \draw[flow] (init) -- (U.west);
    \draw[flow] (U) -- (R)
      node[midway,above,note] {missing};
    \draw[flow] (R) -- (Q)
      node[midway,above,note] {repair due};
    \draw[flow] (Q) -- (S)
      node[midway,above,note] {repair sent};
    \draw[flow] (S.south) .. controls (55.5,-14) and (18,-14) .. (U.south)
      node[midway,below,note,yshift=-0.8mm] {suppression expires while missing};
    \draw[confirm] (nonA.north) -- (A.south)
      node[midway,right,note,xshift=0.8mm] {reception\\confirmed};
  \end{tikzpicture}
  }
  \caption{{Writer-side repair state for one descriptor.} Confirmation
  moves any non-\(\mathrm A\) state to \(\mathrm A\).}
  \label{fig:repair-state}
  \label{fig:sedp-repair-mealy}
  \vspace{-1em}
\end{figure}
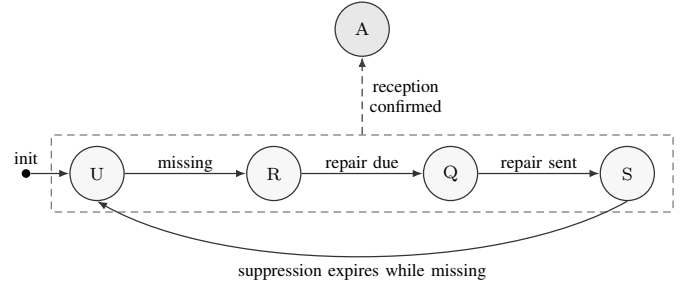

{Fig.~\ref{fig:repair-state} summarizes the sender-side repair state.}

\looseness=-1 {Every descriptor starts in \(\mathrm U\).} An ACKNACK that reports it missing moves it to \(\mathrm R\).
Repair preparation moves it to \(\mathrm Q\).
Completion of repair source transmission moves it to \(\mathrm S\), and suppression expiration returns a still-missing descriptor to \(\mathrm U\).
An ACKNACK that confirms delivery moves any non-\(\mathrm A\) state to \(\mathrm A\).

{A single time-ordered event queue holds message-generation, source-completion, complete-reception, and timer-expiration events.
Each pending entry has the form \((e,t)\), where \(e\) specifies what happens and \(t\) gives its scheduled processing time.
Let \(\mathsf E_n\) denote all entries awaiting processing immediately before step \(n\).
Executing the entry selected at this step may schedule zero or more additional entries, whose set is denoted by \(\Delta\mathsf E_n\).
The queue update is}
\begin{equation}
\begin{aligned}
(e_n,t_n)&=\min_{\prec}\mathsf E_n,\\
\mathsf E_{n+1}&=(\mathsf E_n\setminus\{(e_n,t_n)\})
\cup\Delta\mathsf E_n, 
\end{aligned}
\label{eq:event-evolution}
\end{equation}
{where the pending entry \((e_n,t_n)\) is selected according to the ordering \(\prec\).
After it is removed from \(\mathsf E_n\), every newly scheduled entry in \(\Delta\mathsf E_n\) is inserted to form \(\mathsf E_{n+1}\).
The ordering gives priority to the earliest timestamp.
If message reception completes exactly when the protocol is scheduled to check delivery or decide on repair, reception updates receiver state before that decision is made.
Wireless service begins after every event scheduled for that time has been processed.
Any remaining simultaneous events follow their insertion order.} {Five protocol timers determine when announcements, replies, repairs, and repeated delivery checks may occur:
\begin{equation*}
\begin{aligned}
\tau_A &: \text{HEARTBEAT response delay},\\
\tau_N &: \text{negative acknowledgment~(NACK) response delay},\\
\tau_S &: \text{NACK suppression duration},\\
\tau_H &: \text{HEARTBEAT period},\\
\tau_P &: \text{SPDP resend period}.
\end{aligned}
\end{equation*}
These configured delays and the same-time sequence yield one deterministic trajectory for a fixed deployment, DDS profile, and wireless-service profile.}

{Six rules close the feedback loop.
Here, \(\mathrm{DATA}\) denotes SEDP DATA, \(\mathrm{gen}\) creates a message at its sender, and \(\mathrm{exp}\) marks timer expiration.
The direction \(j\to i\) identifies an ACKNACK sender and receiver.
Finally, \(\mathrm{SUP}_{ijs}\) denotes suppression expiration: it ends the wait after Repair SEDP DATA is sent, during which another repair for the same descriptor is not generated.
The ACKNACK bitmap
\[
\mathbf r^{(s)}_{ij}(t)=\{d:X^{(s)}_{ij,d}(t)=0\}.
\]
lists the endpoint metadata that \(j\) has not processed at time \(t\).
An empty bitmap reports no missing metadata.
\begin{itemize}
\setlength{\itemsep}{1pt}
\setlength{\parskip}{0pt}
\setlength{\parsep}{0pt}
\item \emph{SPDP reception.} When participant \(j\) completely receives
participant \(i\)'s SPDP announcement at time \(t\), the model sets
\(P_{ij}=1\) and \(R^{(s)}_{ij}=1\) for
\(s\in\{\mathrm{pub},\mathrm{sub}\}\), then adds the relation-opening events
specified by the DDS profile.
\item \emph{HEARTBEAT-timer expiration.} When
\((\mathrm{HB}^{\mathrm{exp}},t)\) is processed for participant \(i\)'s
type-\(s\) reliable writer, the following events are added if at least one
descriptor remains unacknowledged:
\[
(\mathrm{HB}^{\mathrm{gen}},t),\quad
(\mathrm{HB}^{\mathrm{exp}},t+\tau_H)
\in\Delta\mathsf E_n .
\]
Neither event is added if all its descriptors~are~\mbox{acknowledged.}
\item \emph{HEARTBEAT reception.} When participant \(j\) completely receives
a HEARTBEAT from participant \(i\)'s type-\(s\) SEDP writer at time \(t\), the
model adds \((\mathrm{ACK}^{\mathrm{gen}}_{j\to i,s},t+\tau_A)\). At its
scheduled time, this event stores
\(\mathbf r^{(s)}_{ij}(t+\tau_A)\) in the ACKNACK. Once generated, its bitmap
does not change.
\item \emph{ACKNACK reception.} When participant \(i\)'s type-\(s\) SEDP
writer receives participant \(j\)'s ACKNACK with a nonempty bitmap at time
\(t\), each newly requested descriptor changes as
\(C^{(s)}_{ij,d}:\mathrm U\rightarrow\mathrm R\), and the model adds
\((\mathrm{DATA}^{\mathrm{gen}},t+\tau_N)\). That event moves the requested
descriptors from \(\mathrm R\) to \(\mathrm Q\) and generates their Repair
SEDP DATA, which enters the backlog and wireless service used by Initial SEDP
DATA.
\item \emph{Repair delivery.} Completion of the final uplink fragment of
Repair SEDP DATA sent from participant \(i\) to participant \(j\) at time
\(t\) changes \(C^{(s)}_{ij,d}:\mathrm Q\rightarrow\mathrm S\) and adds the
suppression-expiration event
\((\mathrm{SUP}_{ijs},t+\tau_S)\). Complete reception by participant \(j\)
changes
\(X^{(s)}_{ij,d}:0\rightarrow1\).
\item \emph{Post-repair wait expiration.} When \(\mathrm{SUP}_{ijs}\) occurs
at time \(t\), every still-unacknowledged descriptor returns as
\(C^{(s)}_{ij,d}:\mathrm S\rightarrow\mathrm U\). If at least one descriptor
returns to \(\mathrm U\), \((\mathrm{HB}^{\mathrm{exp}},t+\tau_H)\) is added
to check its delivery again.
\end{itemize}
}

{Only complete receiver delivery updates \(P\), \(R\), or \(X\), so an earlier timer expiration can generate traffic from the preceding receiver state.}

For example, consider an Initial SEDP DATA message carrying descriptor \(d\).
If it remains queued when a HEARTBEAT reaches participant \(j\), then \(X^{(s)}_{ij,d}=0\), and \(j\)'s ACKNACK snapshots \(d\) as missing. {The bitmap does not change after the ACKNACK is generated.} Even if the initial DATA subsequently reaches \(j\), {the ACKNACK generated before that delivery} can still cause participant \(i\) to prepare Repair SEDP DATA.
That repair reenters the same backlog and competes with the remaining descriptors, potentially delaying more receiver updates beyond later timer expirations.
A later clean ACKNACK moves \(C^{(s)}_{ij,d}\) to \(\mathrm A\) and ends this repair path.

\subsection{Model Outputs}
\label{subsec:message-accounting}

Discovery finishes when every required receiver has processed {every descriptor advertised by a participant on a different host}:
\begin{equation}
\begin{aligned}
\operatorname{Complete}(t)
&\equiv
\bigwedge_{\substack{(i,j)\in\mathcal R\\
s\in\{\mathrm{pub},\mathrm{sub}\}\\
0\le d<D_{i,s}}}
\left(X^{(s)}_{ij,d}(t)=1\right),\\
T_{\mathrm{disc}}
&=\min\{t:\operatorname{Complete}(t)\}.
\end{aligned}
\label{eq:strict-completion}
\end{equation}

{Before \(T_{\mathrm{disc}}\), the model counts each generated logical message in \(M_{\mathrm{init}}\), \(M_{\mathrm{repair}}\), \(M_{\mathrm{HB}}\), \(M_{\mathrm{ACK}}\), \(M_{\mathrm{SPDP,m}}\), or \(M_{\mathrm{SPDP,u}}\), according to its message category.} The first generated occurrence of descriptor \((i,j,s,d)\) is counted as initial DATA, and every later occurrence is counted as repair DATA. {The total message count is the sum of these six categories.} {A message is counted when it is generated, even if it remains queued or in transmission at \(T_{\mathrm{disc}}\).}

\section{Experimental Evaluation}
\subsection{Experimental Methodology}
\label{subsec:experimental-methodology}

\subsubsection{Testbed and Wireless Configuration}

The evaluation instantiates the wireless-service component with two identical hosts connected through an infrastructure IEEE 802.11n link.
This two-host setup holds the physical deployment fixed while varying participant and endpoint density, revealing how logical discovery load activates feedback over the shared medium.
Wi-Fi provides a controlled shared-medium setting in which discovery uplink and AP downlink traffic contend for the same service resource.
The evaluation profile uses a multicast rate of \(6\) Mbps and a unicast rate of approximately \(144.4\) Mbps.
A separate master {host} coordinates run start times, while packet capture and discovery execution occur on the two wireless hosts.

The evaluated Wi-Fi profile specifies the uplink--downlink route, channel-access overheads, transmission rates, and fragment-service order.
The receiver state, reliability timers, and message-generation rules in Section~\ref{sec:model} remain applicable when a different wireless technology is adopted.
Such a network requires its own route, admission, scheduling, and service parameters in place of the infrastructure Wi-Fi profile.
Experimental validation of these alternative service profiles, including mesh and cellular robotic networks, remains as future work.

\subsubsection{DDS Instantiation}
\label{subsec:implementation-profile}

{This subsection instantiates the general model inputs with the ROS~2 and Fast DDS configuration used in the experiments.
Table~\ref{tab:evaluation-profile} first summarizes the resulting evaluation profile.
We then explain how its entries determine the discovery workload and protocol events.}

\begin{table}[!t]
\caption{Evaluation Profile Supplied to the Model}
\label{tab:evaluation-profile}
\vspace{-1em}
\centering
\footnotesize
\setlength{\tabcolsep}{3pt}
\setlength{\aboverulesep}{0.5ex}
\setlength{\belowrulesep}{0.5ex}
\renewcommand{\arraystretch}{1.08}
\begin{tabular}{@{}>{\raggedright\arraybackslash}m{0.30\columnwidth}
                >{\RaggedRight\arraybackslash}m{0.62\columnwidth}@{}}
\toprule
Component & Evaluated value or behavior \\
\midrule
Wireless service & IEEE 802.11n, \(R_{\mathrm m}=6\) Mbps,
\(R_{\mathrm u}\approx144.4\) Mbps \\
\midrule
ROS~2 descriptors &
\(D_{\mathrm{pub}}^{\mathrm{ROS}}=9\) and
\(D_{\mathrm{sub}}^{\mathrm{ROS}}=8\) per default node, then one
\(460\)-B descriptor per user endpoint \\
\midrule
SPDP schedule & Immediate transmission, five \(100\)-ms startup callbacks,
then period \(\tau_P=3\) s \\
\midrule
Fast DDS SEDP timers &
Initial ACKNACK delay \(\tau_I=70\) ms,
\(\tau_H=1\) s, \(\tau_A=10\) \(\mu\)s, and \(\tau_N=100\) \(\mu\)s \\
\midrule
Fast DDS triggers &
First-recognition SPDP resend, relation-open HEARTBEAT,
Initial ACKNACK, and byte-threshold piggyback HEARTBEAT \\
\bottomrule
\end{tabular}
\vspace{-0.5em}
\end{table}

{The experiments use ROS~2 with Fast DDS 2.6.11.
A default ROS~2 node contributes
\[
D_{\mathrm{pub}}^{\mathrm{ROS}}=9,\qquad
D_{\mathrm{sub}}^{\mathrm{ROS}}=8 .
\]
The evaluated workload adds \(N_E\) application endpoints, so the descriptor count used in Section~\ref{subsec:structural-workload} satisfies
\[
D_{\mathrm{pub}}+D_{\mathrm{sub}}
=D_{\mathrm{pub}}^{\mathrm{ROS}}+D_{\mathrm{sub}}^{\mathrm{ROS}}+N_E
=17+N_E .
\]
Each participant advertises this many endpoint descriptions.} {The airtime required to advertise these descriptors depends on their size.
Fast DDS includes each endpoint's topic name, data type, and communication settings in its SEDP DATA, so a UDP packet carrying one default descriptor ranges from \(452\) to \(532\) B.
Each additional application endpoint uses \(460\) B in the evaluated configuration.
The model uses these sizes when Fast DDS groups endpoint descriptions into packets and computes the resulting fragment sizes \(B(c)\).}

The profile also preserves the implementation behaviors that change traffic generation. {Fast DDS sends SPDP announcements by multicast and, after learning another participant's address, also sends a unicast copy directly to that participant.
When an SEDP relation opens at time \(t\), the Fast DDS profile adds \((\mathrm{HB}^{\mathrm{gen}},t)\) and, if needed, \((\mathrm{HB}^{\mathrm{exp}},t+\tau_H)\).} {An Initial ACKNACK is an optional RTPS optimization that Fast DDS implements with an Initial ACKNACK timer.
When participant \(j\) opens a type-\(s\) relation to participant \(i\) at time \(t\), the evaluation model adds \((\mathrm{ACK}^{\mathrm{exp}}_{j\to i,s},t+\tau_I)\), where \(\tau_I=70\) ms.
If no HEARTBEAT has arrived when this event is processed, \(j\) generates an Initial ACKNACK and schedules another attempt with twice the delay.} {RTPS allows one message to carry multiple submessages ~\cite{OMG-RTPS25}, and Fast DDS uses this flexibility to append a piggyback HEARTBEAT to outgoing SEDP DATA when the writer reaches its configured byte threshold.} Repair DATA share this allowance and the same wireless service as initial DATA. {Table~\ref{tab:evaluation-profile} therefore supplies the model with the timer values and optional triggers of the evaluated Fast DDS implementation.}

The model inputs are fixed before the model-to-observation comparison.
Host placement and descriptor counts come from the workload configuration.
Descriptor byte sizes come from the serialized Fast DDS discovery metadata, while the reliability timers and message-generation triggers come from the Fast DDS profile.
The physical-layer rates, fragment route, queue rule, and service order come from the evaluated wireless profile.
Measured completion {times and message counts are then used to evaluate the resulting model predictions.}

\subsubsection{Workload and Synchronized Runs}

The evaluation fixes \(N_H=2\), the smallest cross-host setting, while varying \(N_P\in\{1,\ldots,6\}\) and \(N_E\in\{2,4,\ldots,30\}\), producing 90 participant-endpoint configurations.
Each configuration is repeated 15 times.
Every participant creates approximately half of its user endpoints as publishers and the remainder as subscribers over application-level topics.

Before each run, the hosts synchronize their clocks to the master using \texttt{chronyc}.
The master distributes a target time, and both hosts start packet capture and load the workload before that time.
At the target, each process begins creating its ROS~2 node and configured endpoints, which starts SPDP and SEDP.
Each process polls the ROS middleware~(RMW) matched-endpoint counters every millisecond.
The first instant at which all expected matches are visible is the application-level discovery completion time.

\subsubsection{Trace Processing and Merge}

Each host records its transmitted packets, and the two source-side captures are merged by run so that a packet observed at both hosts is not counted twice.
The application completion time is the observation cutoff.
Because one RTPS/UDP {packet can carry several submessages, we count each contained SPDP, SEDP DATA, HEARTBEAT, and ACKNACK separately.}

SEDP DATA is keyed by source and destination globally unique identifier~(GUID) prefixes, SEDP writer entity, and writer sequence number.
The first occurrence of a key is initial DATA, and later occurrences are repair DATA.
HEARTBEAT and ACKNACK counts are restricted to the reliable SEDP endpoints. Before aggregation, we verify that the merged capture contains at least one SEDP DATA occurrence for every required descriptor key defined in Section~\ref{sec:model}.
The reference itself is checked separately against topology-normalized loopback captures in Subsection~\ref{subsec:completion-scaling}.

\subsection{Burst and Repair Separation}

\begin{figure*}[t]
  \centering
  \begin{tabular}{@{}ccc@{}}
  \multicolumn{1}{c}{\footnotesize Loopback} &
  \multicolumn{1}{c}{\footnotesize Ethernet} &
  \multicolumn{1}{c}{\footnotesize Wi-Fi} \\
  \includegraphics[width=0.315\textwidth]
    {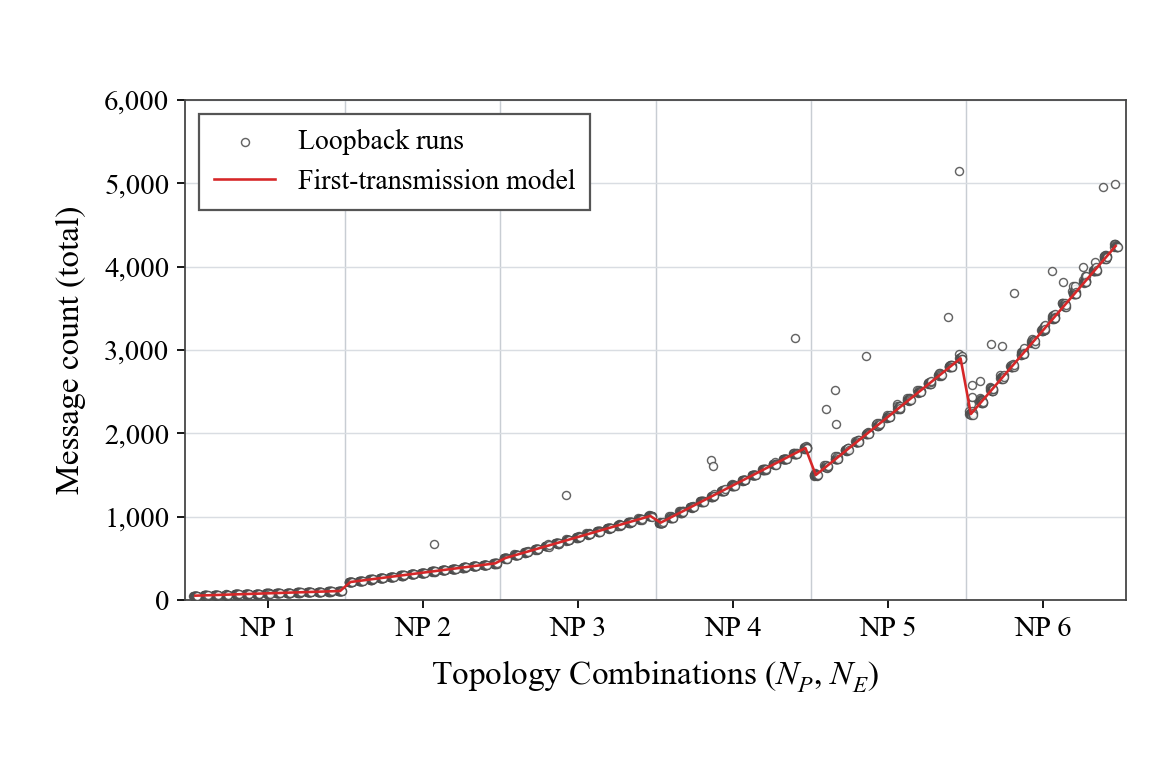} &
  \includegraphics[width=0.315\textwidth]
    {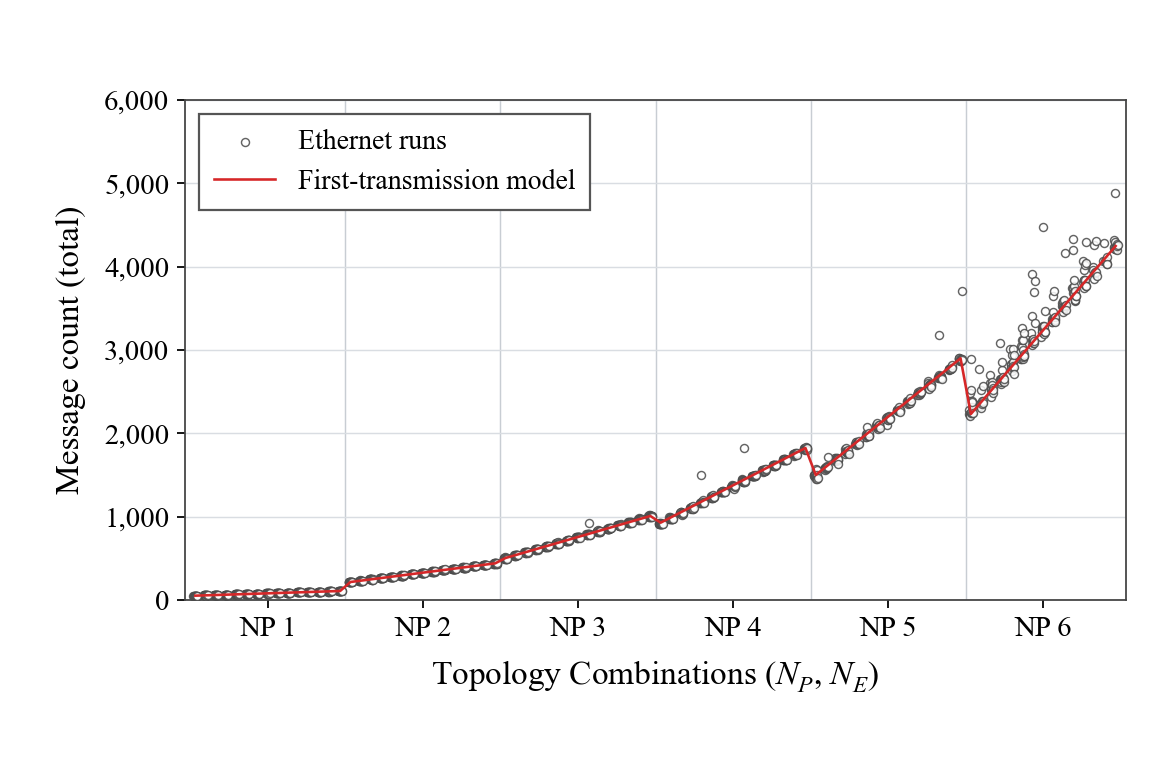} &
  \includegraphics[width=0.315\textwidth]
    {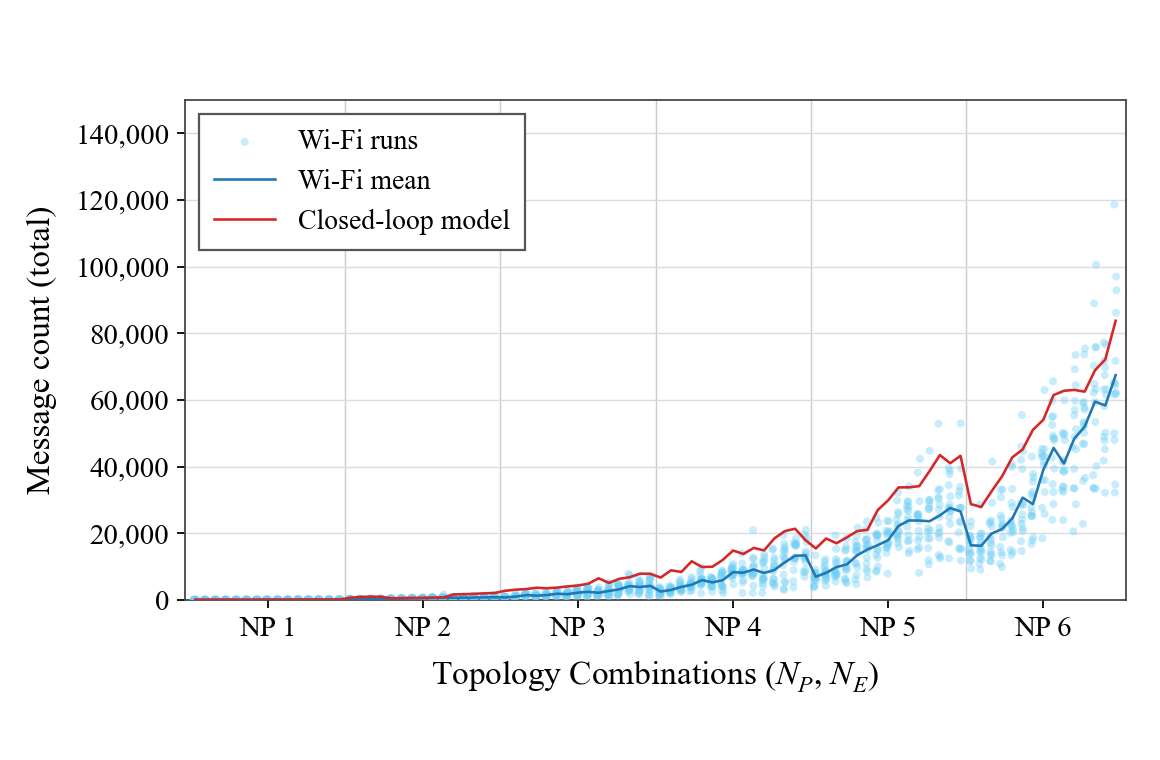} \\
  \includegraphics[width=0.315\textwidth]
    {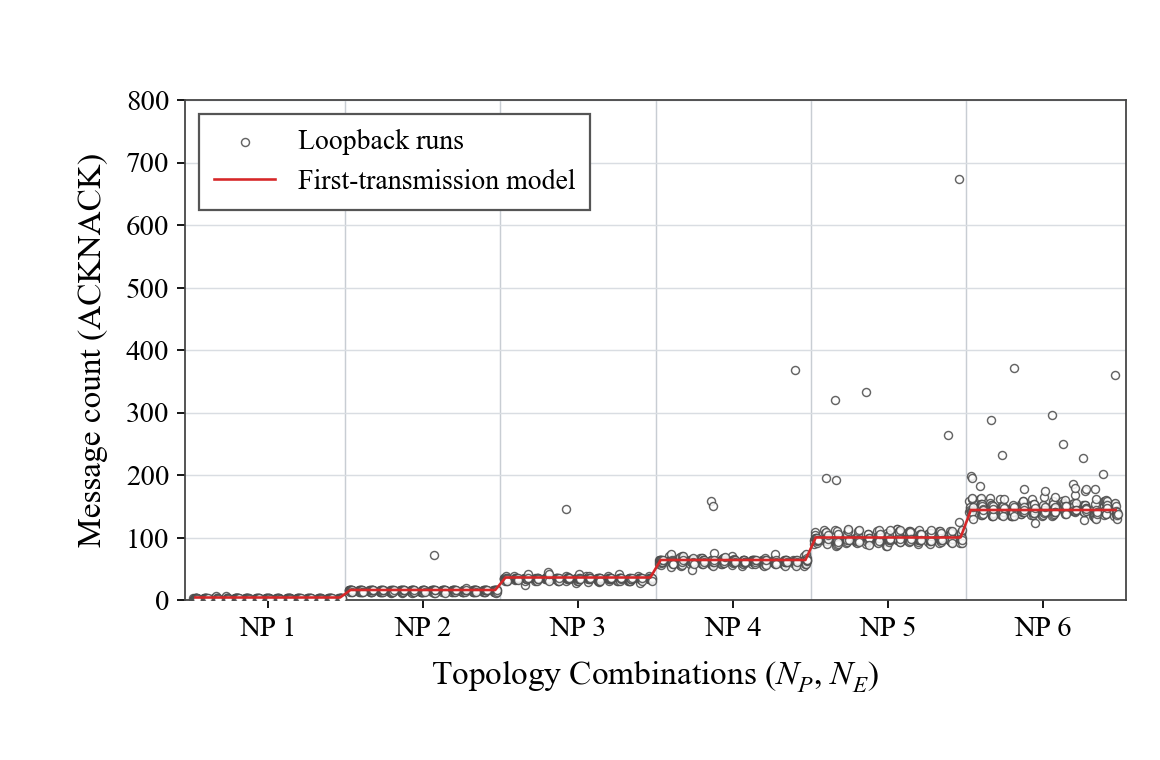} &
  \includegraphics[width=0.315\textwidth]
    {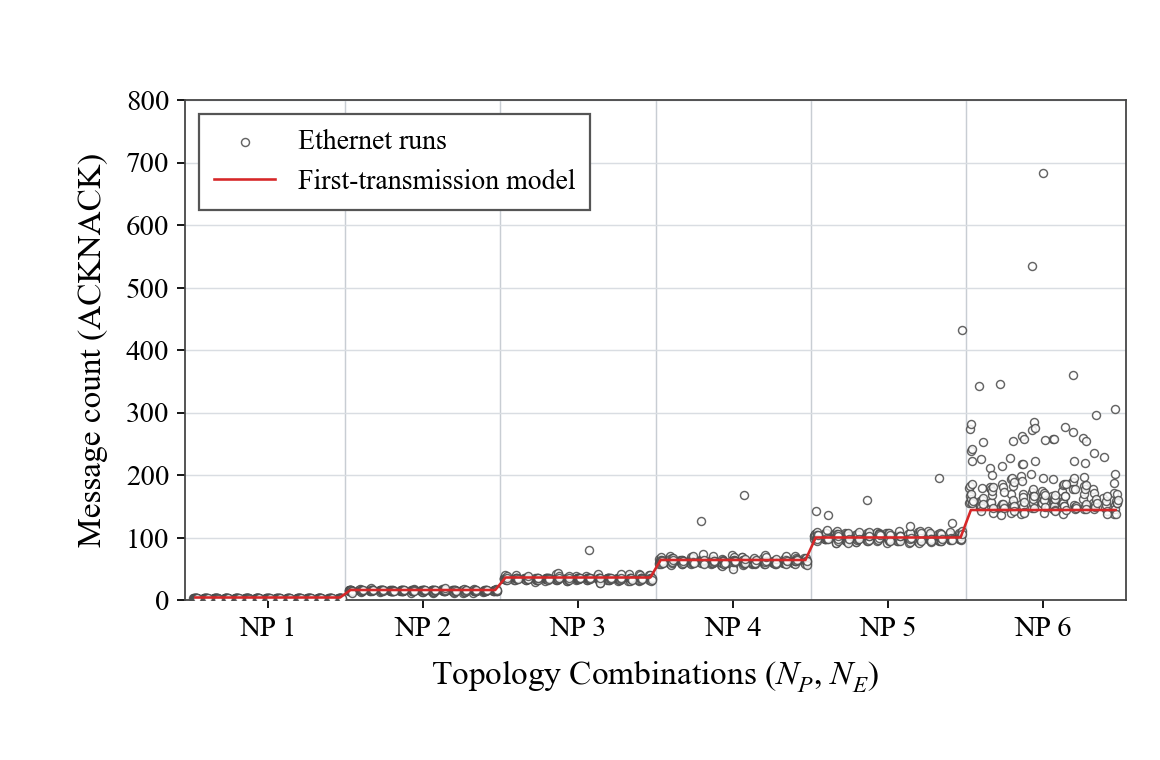} &
  \includegraphics[width=0.315\textwidth]
    {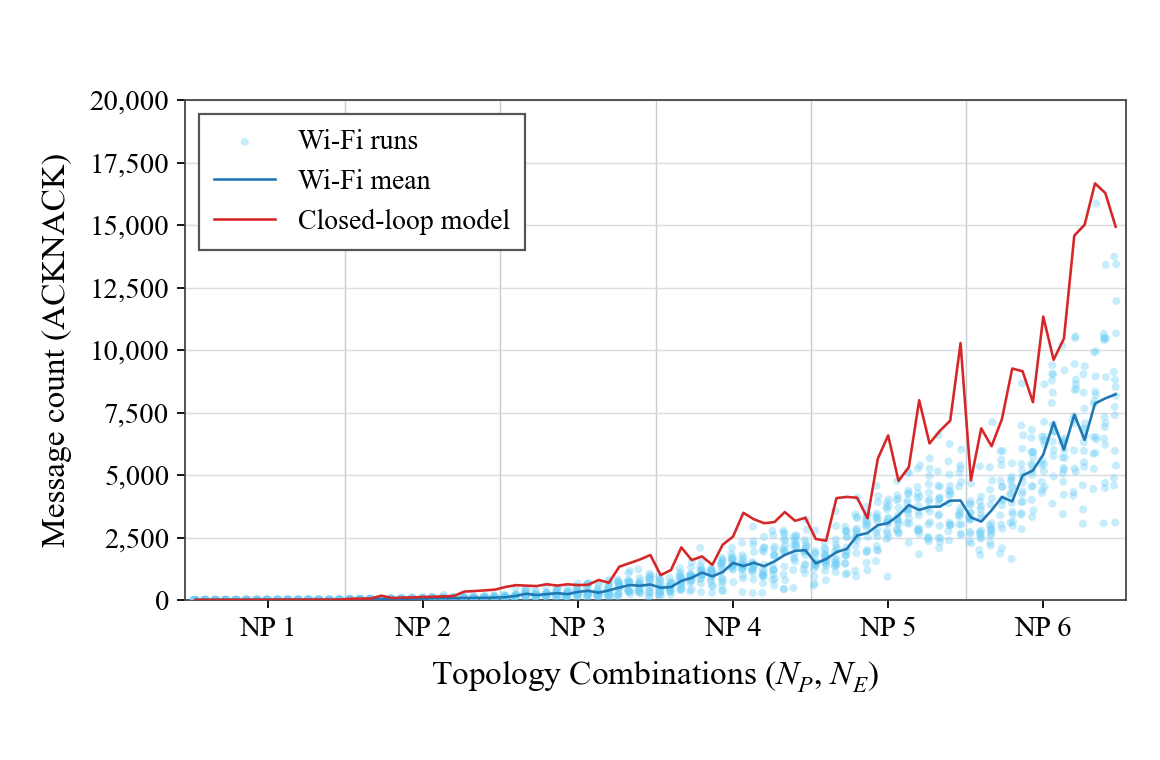}
  \end{tabular}
  \caption{Total discovery submessages (top row) and ACKNACKs
  (bottom row) for loopback, Ethernet, and Wi-Fi. Loopback and Ethernet
  use the same vertical scale in each row. In the Wi-Fi panels, the blue
  line connects configuration means.}
  \label{fig:result-message-wireless}
  \vspace{-1em}
\end{figure*}

{We first separate topology-determined traffic from wireless amplification by comparing loopback, Ethernet, and Wi-Fi over the evaluated configuration range.
Fig.~\ref{fig:result-message-wireless} reports total discovery submessages and ACKNACKs for these three environments.}

{The loopback measurements follow the model's first-transmission reference across all 90 configurations, and Ethernet shows the same failure-free trend.
Their totals therefore provide a failure-free reference for each configuration.
Relative to this reference, Wi-Fi produces substantial increases in both total and ACKNACK counts as delayed receiver recognition generates additional control traffic.
This excess traffic, rather than topology-driven growth alone, is the closed-loop signature of a discovery storm.}

\subsection{Wireless Completion Scaling}
\label{subsec:completion-scaling}

{To evaluate the model's first-transmission delivery reference with negligible network loss, we use loopback captures.
Loopback places all participants on one host, so its number of participant relations differs from that of the evaluated multi-host topology.
Each participant in a topology with \(N_H\) hosts and \(N_P\) participants per host must discover \((N_H-1)N_P\) cross-host peers.
Here, peers are the other participants that must be discovered.
They exclude the participant itself.
Let \(l_P\) be the number of participants in one loopback run.
Matching the peer count gives \(l_P=(N_H-1)N_P+1\).
For the evaluated \(N_H=2\), \(N_P=5\) therefore gives \(l_P=6\).
SEDP DATA, HEARTBEAT, and ACKNACK counts scale with directed participant pairs: \(N_H(N_H-1)N_P^2\) in the multi-host topology and \(l_P(l_P-1)\) in loopback.
Under the evaluated Fast DDS startup profile, SPDP counts scale with the square of the participant population, giving \((N_HN_P)^2\) and \(l_P^2\), respectively.
The ratios give}
\begin{equation}
\begin{aligned}
c_{\mathrm{pair}}
  &=\frac{N_H(N_H-1)N_P^2}{l_P(l_P-1)}
   =\frac{N_HN_P}{(N_H-1)N_P+1},\\
c_{\mathrm{SPDP}}
  &=\left(\frac{N_HN_P}{l_P}\right)^2
   =\frac{N_H^2N_P^2}{\bigl\{(N_H-1)N_P+1\bigr\}^2}.
\end{aligned}
\label{eq:loopback-correction}
\end{equation}
Initial and Repair SEDP DATA, HEARTBEAT, and ACKNACK counts are multiplied by \(c_{\mathrm{pair}}\), while SPDP multicast and unicast counts are multiplied by \(c_{\mathrm{SPDP}}\).
Loopback runs contain {some Repair SEDP DATA in 48 configurations, showing that retransmissions can occur even without wireless delivery.}

\begin{figure}[!t]
  \centering
  \includegraphics[width=0.97\columnwidth]{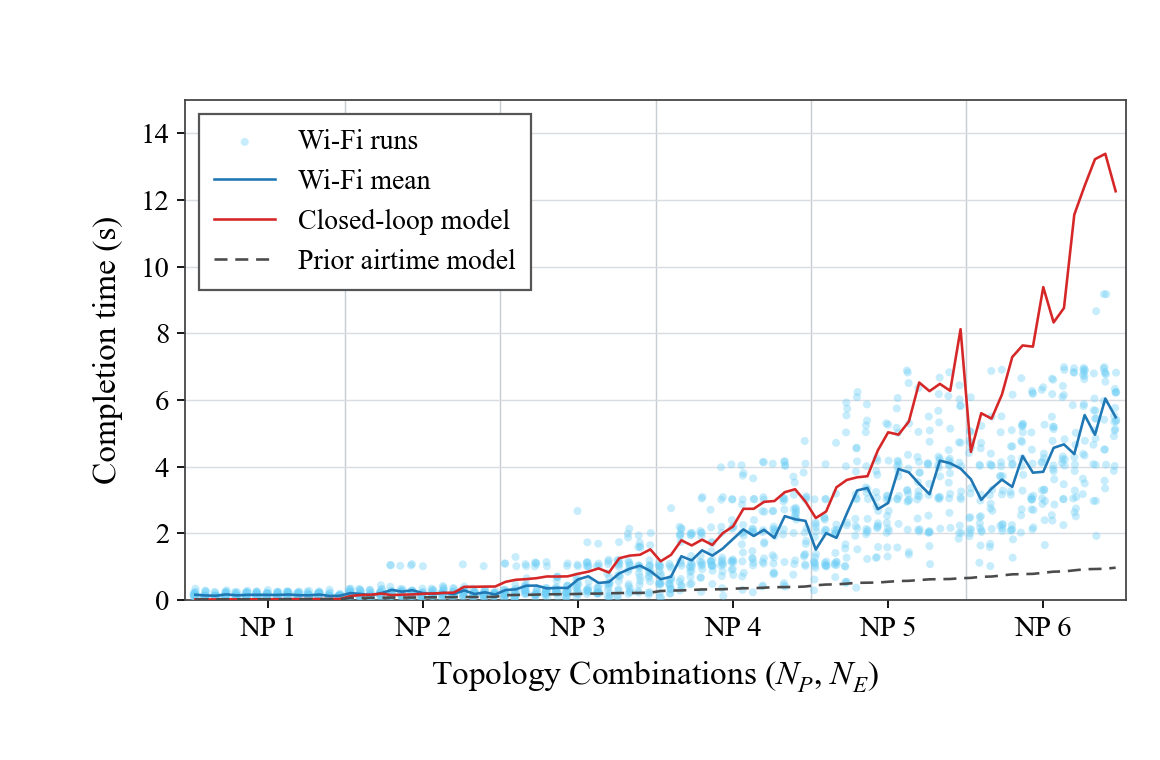}
  \caption{Wireless discovery completion versus workload. Points show
  application proxy-match timestamps, and blue connects configuration means.
  The gray dashed line adapts the open-loop airtime model~\cite{lee2026discovery}
  to the fixed first-transmission traffic, while red shows closed-loop SEDP
  first-recognition completion.}
  \label{fig:result-completion}
  \vspace{-0.5em}
\end{figure}

{Fig.~\ref{fig:result-completion} compares application-level wireless discovery completion with the open-loop airtime and closed-loop models under the Section~III zero middleware-to-application propagation-delay assumption.
Across the 18 high-load configurations, the open-loop airtime model predicts a median 17.1\% of the observed topology mean (IQR 15.7\%--18.4\%), missing the sharp increase.
The closed-loop model captures this increase through feedback from delayed recognition to additional check and repair traffic.
}

\subsection{Model Fit by Message Class}
\label{subsec:model-fit-by-class}

{For the failure-free reference, the topology-normalized loopback measurements have a mean absolute percentage error (MAPE) of \(0.82\%\) for total submessages.
The message-class MAPEs are \(0.33\%\), \(0.68\%\), \(4.14\%\), and \(8.17\%\) for SPDP multicast, SPDP unicast, HEARTBEAT, and ACKNACK, respectively.}

{Table~\ref{tab:upper-coverage} reports the model's upper-range behavior for the wireless runs.
Let \(O\) denote the value observed in one run and \(M\) the corresponding model prediction.
Coverage is the fraction of runs satisfying \(O\leq M\).
For covered runs, the model-to-observation ratio \(M/O\) quantifies the remaining margin, with 1 denoting equality.}

\begin{table}[!t]
\caption{Empirical Upper-Range Behavior by Participant Count}
\label{tab:upper-coverage}
\centering
\scriptsize
\setlength{\tabcolsep}{1.5pt}
\renewcommand{\arraystretch}{1.00}
\begin{tabular*}{\columnwidth}{@{\hspace{6pt}\extracolsep{\fill}}lrrlrr@{\hspace{6pt}}}
\toprule
Scope & Top. & Runs & Quantity & Cov. & Med.\ \(M/O\) [IQR] \\
\midrule
\multirow{2}{*}{\footnotesize \(N_P=1\)} &
\multirow{2}{*}{\footnotesize 15} &
\multirow{2}{*}{\footnotesize 225} &
Completion & 5.8\% & 1.35 [1.28, 1.40] \\
& & & Messages & 28.4\% & 1.34 [1.14, 1.36] \\
\cmidrule(lr){1-6}
\multirow{2}{*}{\footnotesize \(N_P=2\)} &
\multirow{2}{*}{\footnotesize 15} &
\multirow{2}{*}{\footnotesize 225} &
Completion & 64.4\% & 1.51 [1.26, 2.66] \\
& & & Messages & 94.2\% & 2.24 [1.61, 2.98] \\
\cmidrule(lr){1-6}
\multirow{2}{*}{\footnotesize \(N_P=3\)} &
\multirow{2}{*}{\footnotesize 15} &
\multirow{2}{*}{\footnotesize 225} &
Completion & 80.9\% & 2.70 [1.72, 3.94] \\
& & & Messages & 99.1\% & 2.47 [1.77, 3.29] \\
\cmidrule(lr){1-6}
\multirow{2}{*}{\footnotesize \(N_P=4\)} &
\multirow{2}{*}{\footnotesize 15} &
\multirow{2}{*}{\footnotesize 225} &
Completion & 75.1\% & 1.75 [1.38, 2.71] \\
& & & Messages & 97.8\% & 1.98 [1.57, 2.68] \\
\cmidrule(lr){1-6}
\multirow{2}{*}{\footnotesize \(N_P=5\)} &
\multirow{2}{*}{\footnotesize 15} &
\multirow{2}{*}{\footnotesize 225} &
Completion & 85.8\% & 1.77 [1.40, 2.50] \\
& & & Messages & 95.6\% & 1.74 [1.38, 2.20] \\
\cmidrule(lr){1-6}
\multirow{2}{*}{\footnotesize \(N_P=6\)} &
\multirow{2}{*}{\footnotesize 15} &
\multirow{2}{*}{\footnotesize 225} &
Completion & 96.0\% & 2.14 [1.68, 2.84] \\
& & & Messages & 90.2\% & 1.58 [1.27, 1.98] \\
\midrule
\multirow{2}{*}{\footnotesize All} &
\multirow{2}{*}{\footnotesize 90} &
\multirow{2}{*}{\footnotesize 1,350} &
Completion & 68.0\% & 1.96 [1.44, 2.93] \\
& & & Messages & 84.2\% & 1.85 [1.40, 2.63] \\
\cmidrule(lr){1-6}
\multirow{2}{*}{\footnotesize\textbf{High load}} &
\multirow{2}{*}{\footnotesize\textbf{18}} &
\multirow{2}{*}{\footnotesize\textbf{270}} &
\textbf{Completion} &
\textbf{90.7\%} & \textbf{1.98 [1.52, 2.73]} \\
& & & \textbf{Messages} & \textbf{90.4\%} &
\textbf{1.56 [1.29, 2.07]} \\
\bottomrule
\end{tabular*}
\par\vspace{0.3em}
\parbox{\columnwidth}{\scriptsize Each \(N_P\) scope contains 225 runs over
15 values of \(N_E\). High load denotes \(N_P\geq4\) and \(N_E\geq20\).
Coverage is the fraction satisfying \(O\leq M\). Med.\ \(M/O\) [IQR] gives
the median and interquartile range among covered runs.}
\end{table}

\looseness=-1 At low load, omitted middleware processing, polling, and contention-independent losses can dominate.
Shared-airtime feedback instead dominates the targeted high-load region {(\(N_P\geq4\) and \(N_E\geq20\))}.
Table~\ref{tab:upper-coverage} shows respective coverage of \(90.7\%\) and \(90.4\%\) for completion time and total messages, with median covered-run \(M/O\) ratios of 1.98 and 1.56.
Instantaneous startup and per-fragment overhead can contribute to this upper-range tendency.
In experiments, endpoint creation is staggered under OS scheduling, and 802.11n A-MPDU aggregation can reduce effective per-frame overhead.
The model therefore provides conservative upper-range estimates in the storm regime.

Repair SEDP DATA is the most sensitive metric because packet capture cannot identify whether each duplicate was caused by radio-frequency~(RF) failure, queue delay, reordering, or late receiver processing.
We therefore report retransmission fit as diagnostic evidence, not as a direct estimate of physical packet loss.

{We evaluate how the AP decides which transmission goes first when the sender is ready to send the next fragment of a packet at the same time that the AP is ready to forward the preceding fragment to the receiver.
FIFO-Uplink chooses the sender-to-AP transmission, while FIFO-Downlink chooses the AP-to-receiver transmission.
Under Datagram-Uplink, the AP receives all uplink fragments of a packet before forwarding any of them on the downlink.
Across 90 configurations, the median and maximum completion-time spreads are \(0.43\%\) and \(5.53\%\).
Total messages vary by at most \(1.95\%\), and each recovery-message class varies by at most \(3.26\%\).
The small spreads show that the reported trends are insensitive to the choice among the three tested ordering policies.}

\subsection{Model-Guided HEARTBEAT Adaptation}

{The message-class decomposition identifies periodic HEARTBEAT timing as a tunable source of wireless contention.
Because the default interval does not reflect reader responses, HEARTBEATs can contend with pending SEDP DATA during a storm.
We test a response-aware schedule to mitigate this effect, consistent with the broader observation that periodic network processes can synchronize and concentrate shared load~\cite{floyd1994synchronization}.}

{The adaptation changes only periodic HEARTBEATs from the Fast DDS stateful SEDP writers, retaining the default relation-open, ACKNACK-triggered, and piggyback paths.
At each cycle, the writer sends a periodic HEARTBEAT to the set \(\mathcal U\) of readers that have not acknowledged all advertised DATA and records its highest announced sequence number \(q\).
A reader is responsive if its ACKNACK arrives within period \(T\) with an empty bitmap and a base greater than \(q\), indicating no missing DATA through \(q\).
Let \(H=|\mathcal U|\) count the contacted readers and \(A\) the responsive readers.
We then set}
\[
\lambda=\max\!\left\{\frac{A}{H},\epsilon\right\},\qquad
T^\star=\operatorname{clip}\!\left(\frac{T}{\lambda},
T_{\min},T_{\max}\right).
\]
{Each writer starts with \(T=1\) s.
Table~\ref{tab:evaluation-profile} provides \((T_{\min},T_{\max})=(0.1,3.0)\) s, and \(\epsilon=0.05\) keeps \(\lambda\) positive when \(A=0\).
If \(A=H\) before \(T\) expires, the next period becomes the elapsed response time.
If \(T\) expires with \(A<H\), the cycle extends to \(T^\star\), pacing HEARTBEATs to unconfirmed readers in batches of \(\lceil H\lambda\rceil\) until \(T^\star\).
The period returns to 1 s when no unacknowledged change remains.}

\begin{figure}[!t]
  \centering
  \includegraphics[width=0.96\columnwidth]{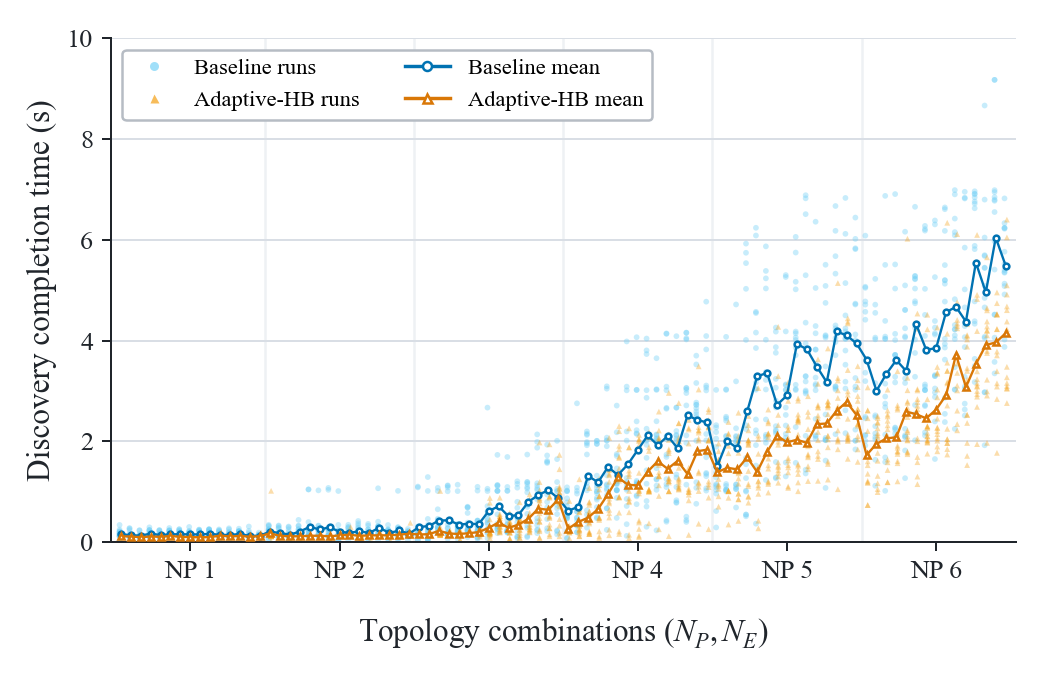}
  \caption{Discovery completion with the default and adaptive HEARTBEAT
  implementations.  Sky-blue circles and orange triangles denote individual
  baseline and adaptive runs, respectively. The blue and orange lines connect
  the corresponding per-configuration means.}
  \label{fig:adaptive-hb-completion}
\end{figure}

{We compare default and adaptive Fast DDS over the same 90 configurations in Subsection~\ref{subsec:experimental-methodology}.
This leaves 1,350 observations each for the baseline and adaptive configurations.
Fig.~\ref{fig:adaptive-hb-completion} shows their individual completion times and configuration means, with mean reductions of \(25.3\%\) to \(39.7\%\) across participant counts.}

{Table~\ref{tab:adaptive-hb-messages} equally weights the 15 endpoint-count means for each \(N_P\) after averaging runs by configuration.
At \(N_P\geq4\), total-message counts rise by \(9.2\%\) to \(16.8\%\), while completion time falls by \(31.5\%\) to \(36.3\%\).
Thus, the schedule improves completion by changing when traffic contends with pending discovery DATA, not by consistently reducing its volume.}

\begin{table}[!t]
\caption{Topology-Balanced Adaptation Summary by \(N_P\).}
\label{tab:adaptive-hb-messages}
\vspace{-1em}
\centering
\footnotesize
\setlength{\tabcolsep}{1.5pt}
\renewcommand{\arraystretch}{1.05}
\begin{tabular*}{\columnwidth}{@{\extracolsep{\fill}}r r r r r@{}}
\toprule
& \multicolumn{2}{c}{Completion} &
\multicolumn{2}{c}{Total messages} \\
\(N_P\) & B/A (s) & \(\Delta\) (\%) & B/A & \(\Delta\) (\%) \\
\midrule
1 & 0.147/0.110 & -25.3 & 97.7/97.4 & -0.4 \\
2 & 0.216/0.135 & -37.8 & 537.4/531.5 & -1.1 \\
3 & 0.571/0.344 & -39.7 & 2315.5/2217.0 & -4.3 \\
4 & 1.689/1.158 & -31.5 & 7466.1/8152.8 & +9.2 \\
5 & 3.127/1.991 & -36.3 & 18116.0/21153.4 & +16.8 \\
6 & 4.307/2.886 & -33.0 & 37939.7/42544.4 & +12.1 \\
\bottomrule
\end{tabular*}
{\parbox{\columnwidth}{\scriptsize\vspace{2pt}
\(B\) and \(A\) denote the baseline and adaptive means, respectively.}}
\vspace{-0.5em}
\end{table}

\section{Related Work}

\emph{Discovery scaling and reduction.} Prior work quantifies the first-transmission SPDP and SEDP burst that precedes a discovery storm~\cite{lee2026discovery}, whereas our model captures the closed-loop feedback that can stall discovery even between two hosts.
Wireless broadcast-storm research instead attributes overload to redundant rebroadcasts, contention, and collisions~\cite{ni2002broadcast}.
DDS-specific techniques reduce initial demand through metadata compression~\cite{sanchez2011bloom}, content-based filtering~\cite{an2014content}, centralized discovery ~\cite{luo2025centralized}, node composition~\cite{macenski2023impact}, or a compact representation~\cite{nwadiugwu2023mad}.
Across these studies, the primary quantities are initial metadata demand and redundant dissemination traffic.

\emph{DDS over wireless robotic networks.} ROS~2 and DDS have been benchmarked across middleware implementations ~\cite{maruyama2016exploring,bode2023systematic} and multi-node latency paths ~\cite{kronauer2021latency}.
DDS has also been studied under multicast load ~\cite{peeroo2022exploring}, industrial wireless channels ~\cite{almadani2015performance}, lossy autonomous links ~\cite{thulasiraman2022evaluation}, and ROS~2 swarm deployments ~\cite{castillo2024swarm}.
These measurements establish implementation choice, network conditions, and deployment scale as major determinants of performance.

\emph{Reliability modeling and network adaptation.} Formal DDS timing analysis bounds data delivery for Fast DDS~\cite{sciangula2023bounding}, while analytical reliability models use an externally specified packet-delivery ratio ~\cite{park2025analytical}.
Periodic-message studies show how timers can synchronize shared load~\cite{floyd1994synchronization}, and network-aware ROS resource management adapts communication using programmable-network information ~\cite{szabo2023toward}.
In these models, delivery conditions enter as timing bounds, specified probabilities, or external network information.

\looseness=-1 \emph{Relation to this work.} Our model links shared-channel service to the metadata recognized by each receiver, whose state then governs delivery checks and repairs.
The resulting traffic reenters the shared channel and changes when the remaining discovery messages arrive.
This closed loop connects traffic generation with delivery instead of treating either one as a fixed input.

\section{Conclusions and Future Work}

This work separates the endpoint information required at startup from the feedback that turns delayed delivery into a discovery storm.
The model links shared-channel delivery, receiver-recognized state, and reliability traffic to the resulting metadata delay.
A high density of endpoints and discovery relations can activate this feedback even between a robot computer and its control station.

{Experiments across 90 two-host configurations show that the model tracks configuration-level trends in discovery completion and message overhead while approximating their observed upper range.
These results show how delayed delivery amplifies repair and delivery-check traffic beyond the failure-free reference.}

{Response-aware HEARTBEAT pacing reduces mean completion time by \(25.3\%\) to \(39.7\%\), even when total traffic increases, showing that recovery timing matters in addition to volume.
This adaptation is a first mitigation step.
We will use the model to suppress the feedback loop itself and move wireless discovery toward the failure-free reference.
We will also extend the model and its validation beyond synchronized homogeneous startup to asynchronous joins, other DDS implementations, and additional wireless networks.}

\bibliographystyle{IEEEtran}

\bibliography{references}

@inproceedings{lee2026discovery,
  author    = {Lee, Sanghoon and Choi, Yeonwoo and Chae, Jiyeong and Park, Kyung-Joon},
  title     = {{Discovery Storm}: Scalability analysis of {DDS} and {Zenoh} in large-scale wireless robotic networks},
  booktitle = {Proceedings of {IEEE} {INFOCOM}},
  pages     = {1--6},
  month     = may,
  year      = {2026},
  doi       = {10.1109/INFOCOM59046.2026.11571268}
}

@article{macenski2022robot,
  author  = {Macenski, Steve and Foote, Tully and Gerkey, Brian and Lalancette, Chris and Woodall, William},
  title   = {{Robot Operating System 2}: Design, architecture, and uses in the wild},
  journal = {Science Robotics},
  volume  = {7},
  number  = {66},
  pages   = {eabm6074},
  month   = may,
  year    = {2022},
  doi     = {10.1126/scirobotics.abm6074}
}

@techreport{OMG-RTPS25,
  author      = {{Object Management Group}},
  title       = {{The Real-Time Publish-Subscribe Protocol: DDS Interoperability Wire Protocol (DDSI-RTPS), Version 2.5}},
  institution = {Object Management Group},
  number      = {formal/22-04-01},
  month       = apr,
  year        = {2022},
  url         = {https://www.omg.org/spec/DDSI-RTPS/2.5}
}

@inproceedings{park2025analytical,
  author    = {Park, Hyung-Seok and Lee, Sanghoon and Um, Doosik and Ryu, Hyunho and Park, Kyung-Joon},
  title     = {An analytical latency model of the {Data Distribution Service} in {ROS 2}},
  booktitle = {Proceedings of {IEEE} {INFOCOM}},
  pages     = {1--10},
  month     = may,
  year      = {2025},
  doi       = {10.1109/INFOCOM55648.2025.11044454}
}

@article{sanchez2011bloom,
  author  = {S{\'a}nchez-Monedero, Javier and Povedano-Molina, Javier and L{\'o}pez-Vega, Jos{\'e} M. and L{\'o}pez-Soler, Juan M.},
  title   = {Bloom filter-based discovery protocol for {DDS} middleware},
  journal = {Journal of Parallel and Distributed Computing},
  volume  = {71},
  number  = {10},
  pages   = {1305--1317},
  month   = oct,
  year    = {2011},
  doi     = {10.1016/j.jpdc.2011.05.001}
}

@inproceedings{an2014content,
  author    = {An, Kyoungho and Gokhale, Aniruddha S. and Schmidt, Douglas C. and Tambe, Sumant and Pazandak, Paul and Pardo-Castellote, Gerardo},
  title     = {Content-based filtering discovery protocol ({CFDP}): Scalable and efficient {OMG DDS} discovery protocol},
  booktitle = {Proceedings of the 8th {ACM} International Conference on Distributed Event-Based Systems ({DEBS})},
  pages     = {130--141},
  month     = may,
  year      = {2014},
  doi       = {10.1145/2611286.2611300}
}

@article{luo2025centralized,
  author  = {Luo, Feng and Ren, Yi and Yu, Yanhua and Li, Yunpeng and Liu, Qin and Zhang, Xiaobo},
  title   = {A centralized discovery-based method for integrating {Data Distribution Service} and {Time-Sensitive Networking} for {In-Vehicle Networks}},
  journal = {Ad Hoc Networks},
  volume  = {178},
  pages   = {103950},
  month   = nov,
  year    = {2025},
  doi     = {10.1016/j.adhoc.2025.103950}
}

@article{macenski2023impact,
  author  = {Macenski, Steve and Soragna, Alberto and Carroll, Michael and Ge, Zhenpeng},
  title   = {Impact of {ROS 2} node composition in robotic systems},
  journal = {{IEEE} Robotics and Automation Letters},
  volume  = {8},
  number  = {7},
  pages   = {3996--4003},
  month   = jul,
  year    = {2023},
  doi     = {10.1109/LRA.2023.3279614}
}

@inproceedings{thulasiraman2022evaluation,
  author    = {Thulasiraman, Preetha and Cheng, Yu Kheng Denny and Allen, Bruce},
  title     = {Evaluation of the {Data Distribution Service} for a lossy autonomous hybrid system},
  booktitle = {Proceedings of the 2022 {IEEE} International Systems Conference ({SysCon})},
  pages     = {1--8},
  month     = apr,
  year      = {2022},
  doi       = {10.1109/SysCon53536.2022.9773896}
}

@article{castillo2024swarm,
  author  = {Castillo-S{\'a}nchez, Jos{\'e}-Borja and Gonz{\'a}lez-Parada, Eva and Cano-Garc{\'\i}a, Jos{\'e}-Manuel},
  title   = {Swarm robot communications in {ROS 2}: An experimental study},
  journal = {{IEEE} Access},
  volume  = {12},
  pages   = {142930--142943},
  year    = {2024},
  doi     = {10.1109/ACCESS.2024.3470254}
}

@article{szabo2023toward,
  author  = {Szab{\'o}, G{\'e}za},
  title   = {Toward the automatic network resource management of {Robot Operating System} in programmable mobile networks},
  journal = {{IEEE} Access},
  volume  = {11},
  pages   = {65934--65955},
  year    = {2023},
  doi     = {10.1109/ACCESS.2023.3289922}
}

@article{nwadiugwu2023mad,
  author  = {Nwadiugwu, Williams-Paul and Kim, Dong-Seong and Ejaz, Waleed and Anpalagan, Alagan},
  title   = {{MAD-DDS}: Memory-efficient automatic discovery {Data Distribution Service} for large-scale distributed control network},
  journal = {{IET} Communications},
  volume  = {17},
  number  = {12},
  pages   = {1432--1446},
  month   = jul,
  year    = {2023},
  doi     = {10.1049/cmu2.12645}
}

@inproceedings{maruyama2016exploring,
  author    = {Maruyama, Yuya and Kato, Shinpei and Azumi, Takuya},
  title     = {Exploring the performance of {ROS 2}},
  booktitle = {Proceedings of the 13th {ACM} International Conference on Embedded Software ({EMSOFT})},
  pages     = {1--10},
  month     = oct,
  year      = {2016},
  doi       = {10.1145/2968478.2968502}
}

@inproceedings{kronauer2021latency,
  author    = {Kronauer, Tobias and Pohlmann, Joshwa and Matth{\'e}, Maximilian and Smejkal, Till and Fettweis, Gerhard},
  title     = {Latency analysis of {ROS 2} multi-node systems},
  booktitle = {Proceedings of the 2021 {IEEE} International Conference on Multisensor Fusion and Integration for Intelligent Systems ({MFI})},
  pages     = {1--7},
  month     = sep,
  year      = {2021},
  doi       = {10.1109/MFI52462.2021.9591166}
}

@inproceedings{bode2023systematic,
  author    = {Bode, Vincent and Buettner, David and Preclik, Tobias and Trinitis, Carsten and Schulz, Martin},
  title     = {Systematic analysis of {DDS} implementations},
  booktitle = {Proceedings of the 24th {ACM}/{IFIP} International Middleware Conference},
  pages     = {234--246},
  year      = {2023},
  doi       = {10.1145/3590140.3629118}
}

@inproceedings{sciangula2023bounding,
  author    = {Sciangula, Gerlando and Casini, Daniel and Biondi, Alessandro and Scordino, Claudio and Di Natale, Marco},
  title     = {Bounding the data-delivery latency of {DDS} messages in real-time applications},
  booktitle = {Proceedings of the 35th Euromicro Conference on Real-Time Systems ({ECRTS})},
  series    = {Leibniz International Proceedings in Informatics},
  volume    = {262},
  pages     = {9:1--9:26},
  month     = jul,
  year      = {2023},
  doi       = {10.4230/LIPIcs.ECRTS.2023.9}
}

@inproceedings{peeroo2022exploring,
  author    = {Peeroo, Kaleem and Popov, Peter and Stankovic, Vladimir},
  title     = {Exploring the effects of multicast communication on {DDS} performance},
  booktitle = {Proceedings of the 18th European Dependable Computing Conference ({EDCC}) Student Forum},
  month     = sep,
  year      = {2022},
  doi       = {10.48550/arXiv.2209.09001}
}

@article{almadani2015performance,
  author  = {Almadani, Basem and Bajwa, Muhammad Naseer and Yang, Shuang-Hua and Saif, Abdul-Wahid A.},
  title   = {Performance evaluation of {DDS}-based middleware over wireless channel for reconfigurable manufacturing systems},
  journal = {International Journal of Distributed Sensor Networks},
  volume  = {11},
  number  = {7},
  pages   = {863123},
  month   = jul,
  year    = {2015},
  doi     = {10.1155/2015/863123}
}

@article{floyd1994synchronization,
  author  = {Floyd, Sally and Jacobson, Van},
  title   = {The synchronization of periodic routing messages},
  journal = {{IEEE}/{ACM} Transactions on Networking},
  volume  = {2},
  number  = {2},
  pages   = {122--136},
  month   = apr,
  year    = {1994},
  doi     = {10.1109/90.298431}
}

@article{ni2002broadcast,
  author  = {Ni, Sze-Yao and Tseng, Yu-Chee and Chen, Yuh-Shyan and Sheu, Jang-Ping},
  title   = {The broadcast storm problem in a mobile ad hoc network},
  journal = {Wireless Networks},
  volume  = {8},
  number  = {2--3},
  pages   = {153--167},
  year    = {2002},
  doi     = {10.1023/A:1013763825347}
}

\end{document}